\documentclass[aps,physrev,preprint,superscriptaddress]{revtex4-2}

\usepackage{graphicx}

\makeatletter
\def\@bibdataout@aps{%
 \immediate\write\@bibdataout{%
  @CONTROL{%
   apsrev42Control%
   \longbibliography@sw{%
    ,author="9",editor="0",pages="0",title="0",year="1"%
   }{%
    ,author="9",editor="0",pages="0",title="",year="1"%
   }%
  }%
 }%
 \if@filesw
  \immediate\write\@auxout{\string\citation{apsrev42Control}}%
 \fi
}%
\makeatother

\begin{document}
\raggedbottom
\vfuzz=4pt
\setlength{\parskip}{0.6\baselineskip}


\title{Imaging the Magnetically Driven Reconstruction of the Electronic States in the Antiferromagnetic Topological Insulator \(\mathrm{EuSn_2As_2}\)}


\author{Luka Khizanishvili}
\affiliation{Department of Physics, Applied Physics and Astronomy, Binghamton University, Binghamton, New York 13902, USA}

\author{Erekle Jmukhadze}
\affiliation{Department of Physics, Applied Physics and Astronomy, Binghamton University, Binghamton, New York 13902, USA}

\author{Anika Raisa}
\affiliation{Department of Physics, Applied Physics and Astronomy, Binghamton University, Binghamton, New York 13902, USA}

\author{Divyanshi Sar}
\affiliation{Department of Physics, Applied Physics and Astronomy, Binghamton University, Binghamton, New York 13902, USA}

\author{Mingda Gong}
\affiliation{Department of Physics, Applied Physics and Astronomy, Binghamton University, Binghamton, New York 13902, USA}

\author{Tetiana Romanova}
\affiliation{Institute of Low Temperature and Structure Research, Polish Academy of Sciences, Ok\'{o}lna 2, 50-422 Wroc\l{}aw, Poland}

\author{Dariusz Kaczorowski}
\affiliation{Institute of Low Temperature and Structure Research, Polish Academy of Sciences, Ok\'{o}lna 2, 50-422 Wroc\l{}aw, Poland}

\author{Wei-Cheng Lee}
\affiliation{Department of Physics, Applied Physics and Astronomy, Binghamton University, Binghamton, New York 13902, USA}

\author{Pegor Aynajian}
    \email{Corresponding author: aynajian@binghamton.edu}
\affiliation{Department of Physics, Applied Physics and Astronomy, Binghamton University, Binghamton, New York 13902, USA}



\date{\today}

\begin{abstract}
The realization of the axion insulator phase in magnetic topological insulators is often hindered by crystalline symmetries that protect gapless surface states, even when time-reversal symmetry is broken. Here, we use variable-temperature scanning tunneling microscopy (STM) and spectroscopy (STS), complemented with density functional theory (DFT), to investigate the local electronic structure of the antiferromagnetic (AFM) topological insulator $EuSn_2As_2$ across its N\'{e}el transition at $T_N = 24 \,\mathrm{K}$. On the (001) surface, we observe substantial density of intrinsic Sn vacancies that introduce nanoscale electronic inhomogeneity and p-type doping. Upon cooling below $T_N$, we resolve the emergence of two distinct magnetically driven gaps: a $\sim 100\,\mathrm{meV}$ gap near the Fermi level and a $\sim 50\,\mathrm{meV}$ gap at the ARPES-resolved Dirac point. We attribute the former gap to the AFM Brillouin zone folding and hybridization. The characteristics of the $50\,\mathrm{meV}$ gap point toward the lifting of mirror-symmetry protection by Sn vacancies and mass-gapping the Dirac point, although contributions from AFM-induced folding hybridization cannot be entirely ruled out. Our findings provide real-space evidence for the strong coupling between localized moments and itinerant topological states, highlighting exfoliable $EuSn_2As_2$ as a potential candidate for realizing axion insulator-based devices.
\end{abstract}


\maketitle

\section{Introduction}
\enlargethispage{6pt}
The axion insulator is a distinct topological state of matter characterized by a quantized topological magnetoelectric (TME) response in the bulk. This phenomenon is described by the addition of an axion term, $\mathcal{L}_{\theta} = \frac{e^2}{2\pi h}\,\theta\, \mathbf{E}\cdot\mathbf{B}$, to the electromagnetic action, where the axion field $\theta$ is quantized to $\pi$ \cite{qi2008topological, essin2009magnetoelectric, li2010dynamical, liu2020robust, wang2010equivalent}. In such materials, an applied magnetic field induces an electric polarization, while an applied electric field generates a parallel magnetization \cite{qi2008topological,essin2009magnetoelectric, li2010dynamical, liu2020robust,mogi2017magnetic}. This response is topologically robust as long as the bulk energy gap remains open and specific symmetries are maintained. Beyond its fundamental significance, the quantized TME effect offers a promising platform for next generation technologies, including magnetoelectric memory and ultra-low-power spintronic devices \cite{smejkal2018topological}.

Axion insulators typically emerge from the subtle interplay between long-range magnetic order and topological band inversion \cite{sekine2021axion}. A particularly promising platform for realizing this phase is found in A-type AFM topological insulators, where spins align ferromagnetically within individual layers but couple antiferromagnetically along the stacking direction \cite{mong2010antiferromagnetic}. In these systems, the magnetic order breaks time-reversal symmetry ($\mathcal{T}$) but preserves inversion symmetry ($\mathcal{P}$). When this AFM order develops from a paramagnetic (PM) strong topological insulator without closing the bulk gap, the system acquires a bulk axion phase characterized by a topological index $Z_4 = 2$ \cite{turner2012quantized}.

The manifestation of the axion state at the crystal surface is more subtle. For a (001) termination, the surface naturally breaks the half translational magnetic symmetry ($t_{1/2}$) and therefore the combined symmetry $\mathcal{S} = \mathcal{T}t_{1/2}$ , which typically protects the gapless Dirac states in out-of-plane AFM (AFM-c) topological insulators \cite{fang2013topological, li2019dirac, honma2023antiferromagnetic}. Yet, in systems where spins are oriented within the plane (the AFM-b phase), residual crystalline symmetries, such as the vertical mirror symmetry $M_y$, may continue to protect the gapless Dirac fermions, and thus realizing a fully gapped axion-insulator surface requires lifting these residual crystalline protections \cite{xu2019higher, riberolles2021magnetic, fang2015new, jing2026strain}.

While much of the experimental effort has focused on the $Mn(Bi,Sb)_{2n}Te_{3n+1}$ family of magnetic topological insulators (where pronounced sample-to-sample variation of the existence or absence of the exchange energy gap of the topological surface states has been reported \cite{otrokov2019prediction, hao2019gapless, chen2019topological,garnica2022native, li2019dirac}), the Eu-based 122 systems have recently emerged as a compelling alternative for realizing the axion insulator phase \cite{xu2019higher, regmi2019temperature, riberolles2021magnetic}. Both $EuIn_2As_2$ \cite{xu2019higher, regmi2019temperature} and $EuSn_2As_2$ \cite{li2019dirac} have been theoretically proposed as robust axion insulators. Previous STM studies on $EuIn_2As_2$ identified an inverted bulk gap and partially gapped surface states \cite{gong2022surface}. Yet, atomic surface reconstructions inherent to this system complicates the identification of the surface magnetization and underlying gapping mechanism \cite{gong2022surface}. As shown in Fig.1(a), the existence of a neutral $Sn-Sn$ van-der Waals bonded layers in the $R\bar{3}m \: (No.166)$ structure of $EuSn_2As_2$ (distinct from the $P6_3/mmc \: (No.194)$ structure of $EuIn_2As_2$ \cite{gong2022surface}), renders this system especially well-suited for experimental feasibility and integration into van der Waals heterostructure-based spintronic devices \cite{ma2025layer, tien2025quantum}. To date, no STM investigation of $EuSn_2As_2$ have been reported.

$EuSn_2As_2$ crystallizes in a rhombohedral structure consisting of $[Sn_2As_2]^{2-}$ layers separated by $Eu^{2+}$ planes and exhibits A-type AFM order below $T_N \approx 24 \,\mathrm{K}$, with Eu $4f$ moments oriented predominantly in the ab plane (AFM-b) \cite{arguilla2017eusn, pakhira2021type}. Transport measurements show that $EuSn_2As_2$ behaves as a low-carrier density semimetal with relatively high resistivity and a clear anomaly at $T_N$ (see Supplemental Material Fig.~S2). Band-structure calculations [Figs.1(b)-1(c)] and angle-resolved photoemission spectroscopy (ARPES) \cite{li2019dirac} show $EuSn_2As_2$ to host an inverted bulk band structure and Dirac surface states on the (001) surface, placing it among the leading candidates for an AFM axion insulator. However, ARPES reports no changes in either the topological surface states or the overall electronic band structure across the magnetic transition \cite{li2019dirac}. This lack of significant modification is attributed to weak hybridization between the localized Eu $4f$ moments and the itinerant electronic states that form the topological bands, indicating that the magnetism and the topological electronic structure may be relatively decoupled in this material.

The ARPES work \cite{li2019dirac} further reveals that the electronic band structure in this system is shifted by $\sim 180 \,\mathrm{meV}$ relative to the Fermi energy (as compared to DFT calculations) likely due to intrinsic defects that hole-dope the material system, which is commonly seen in other Eu-122 systems \cite{li2019dirac, gui2019new, soh2019ideal, cao2022giant, sar2026topologically}. More recent structural studies have also found nano-scale defects, such as planar insertion of a distinct $EuSn_1As_2$ phase \cite{levakhova2025emergence}. These intrinsic defects may introduce nanoscale inhomogeneity within the electronic state and a local probe measurement may therefore provide invaluable information on the role of defects and the local interplay between magnetism and band topology across the magnetic transition.

In this work, we use variable-temperature STM \cite{yazdani2016spectroscopic} complemented by DFT calculations to investigate the temperature-evolution of the surface electronic structure of $EuSn_2As_2$ across the AFM transition. We observe atomically resolved topographs of the cleaved Sn-terminated (001) surface that reveal a substantial density of Sn vacancies, introducing nanoscale inhomogeneity and hole-doping the system. Temperature-dependent tunneling spectroscopy reveals the opening of a $100 \,\mathrm{meV}$ gap near the Fermi energy ($E_F$) upon cooling below $T_N$. In addition, we observe the opening of a second gap below $T_N$ in the energy range where the prior ARPES measurements report the Dirac surface states crossing \cite{li2019dirac}. The intrinsic Sn vacancies resolved by STM lead to nanoscale spatial inhomogeneity in the energy positions of both gaps, which can be averaged out in nonlocal probes. Our findings demonstrate strong coupling between Eu $4f$ local moments and the electronic states, establishing $EuSn_2As_2$ as a magnetic topological material and a viable axion insulator candidate.

\section{Results and Discussion}
Figure 1(d) shows a $60 \: nm$ topograph of in-situ cleaved $EuSn_2As_2$. Except for two cases (see Supplemental Material Fig.~S3), the same surface termination was observed after cleaving five samples and examining more than 40 distinct regions separated by hundreds of microns. Since statistically only one type of surface is observed (which indicates cleaving between two identical layers), together with the van der Waals-like bonding between the Sn layers, we attribute the exposed surface to a Sn termination. Indeed, recent work has shown that $EuSn_2As_2$ can be exfoliated down to a few monolayers \cite{ma2025layer, maltsev2026anomalous, arguilla2017eusn} further establishing the van-der Waals bonding in this system, which is different from other Eu-122 systems. From the atomically resolved STM topograph and atomic steps, we extract the in- and out-of-plane lattice constants of $a = 4.2$ \AA{} and $c/3 = 8.8$ \AA{} [Figs.1(e)-1(g)], which are consistent with the reported values \cite{li2021magnetic,pakhira2021type}. As seen in Figs.1(d)-1(f), the surface exhibits dark defects centered on the surface atomic sites. The defect profile, showing an apparent depth of $\sim 0.6$ \AA{} in the topograph [Figs.1(f)-1(g)], is consistent with Sn vacancies. Statistical analysis of large-area STM images yields a vacancy concentration of approximately $(2.2 \pm 0.2)\%$.

 \begin{figure}
 \includegraphics[width=\textwidth]{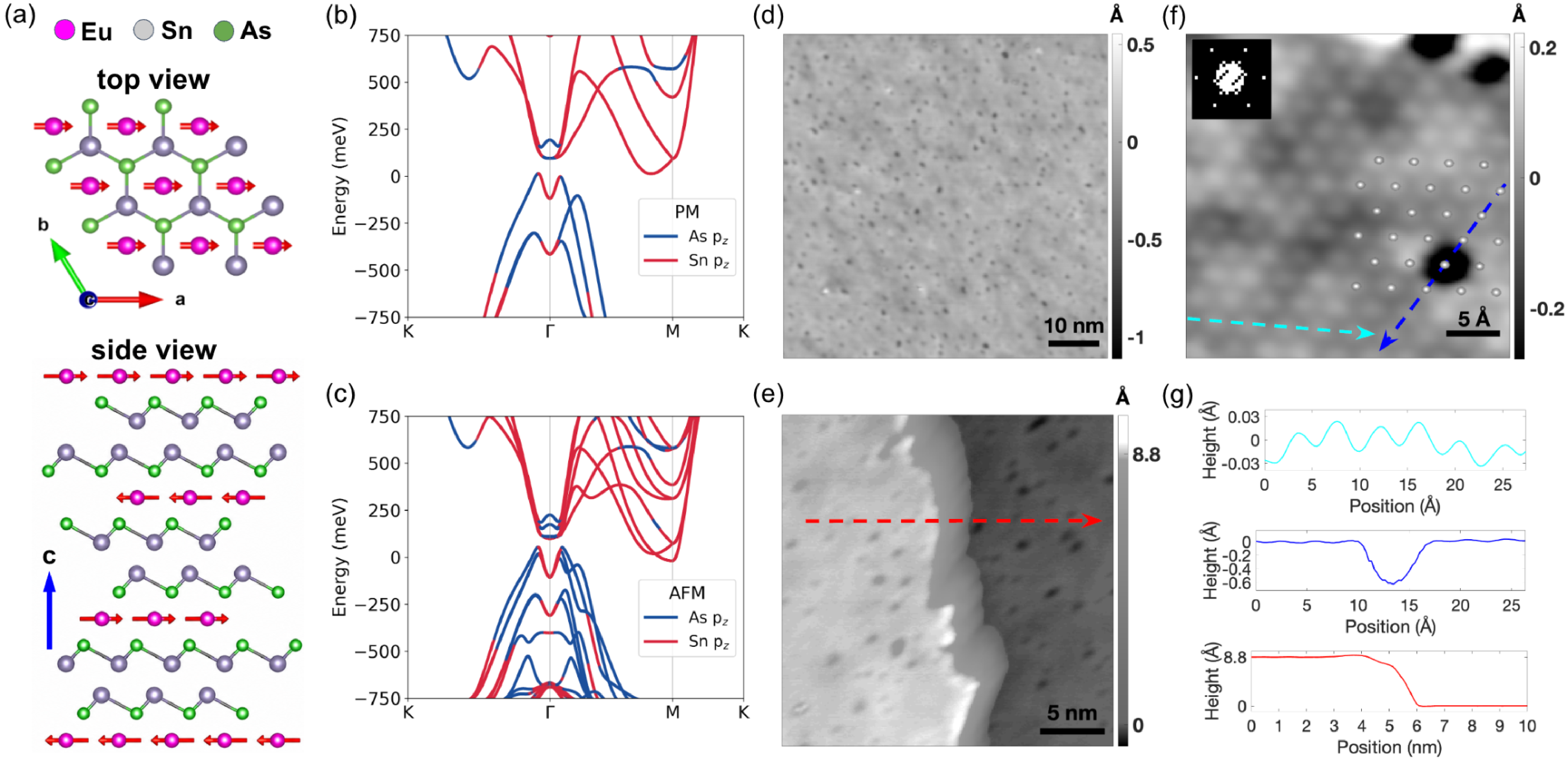}%
 \caption{Crystal structure and electronic topology of $EuSn_2As_2$. (a) Crystal structure of $EuSn_2As_2$. Magenta arrows indicate the spin orientation of Eu moments. (b),(c) Calculated bulk band structures of $EuSn_2As_2$ in the PM and AFM-b phases, respectively, showing the spin-orbit-coupling-induced band gap near the $\Gamma$ point close to $E_F$. (d) Large-area topographic image acquired at $V=-250\,\mathrm{mV}$ and $I_t=-800\,\mathrm{pA}$, displaying a homogeneous distribution of dark defects. (e) Topographic image across a step edge acquired at $V=-400\,\mathrm{mV}$ and $I_t=-800\,\mathrm{pA}$, showing identical surface terminations on both terraces. (f) High-resolution topographic image revealing atomic-scale defects corresponding to missing surface atoms, acquired at $V=-300\,\mathrm{mV}$ and $I_t=-150\,\mathrm{pA}$. (g) Line profiles taken along the cyan and blue arrows in (f), corresponding to the pristine surface and a defect site respectively, and along the red arrow in (e) across the step edge.}
 \end{figure}

The impact of Sn vacancies on the lattice is two-fold. Besides the missing atom in the topographs, the nominal valence of the $Sn^{2+}$ vacancies introduce substantial hole doping that shift the chemical potential. The perturbation induced by such doping in a semi-metallic system extends well beyond the immediate vacancy site. To investigate the impact of Sn vacancies on the local electronic states, in Figs.2(a)-2(b) we plot the differential conductance spectra (dI/dV) recorded along the dashed line and the different locations indicated in the topograph shown in Fig.2(c) measured in the AFM phase at temperature of $9\,\mathrm{K}$. The spectra exhibit several prominent features, most notably a $100 \,\mathrm{meV}$ gap-peak structure located near the Fermi energy $E_F$ and a more pronounced gap characterized by low density of states (DOS) in the $300\text{--}500 \,\mathrm{meV}$ range. Within this large gap, a secondary smaller gap at $\sim400 \,\mathrm{meV}$ can also be seen. The observed spectral features further exhibit significant nanoscale spatial variations.

\begin{figure}
\includegraphics[width=\textwidth]{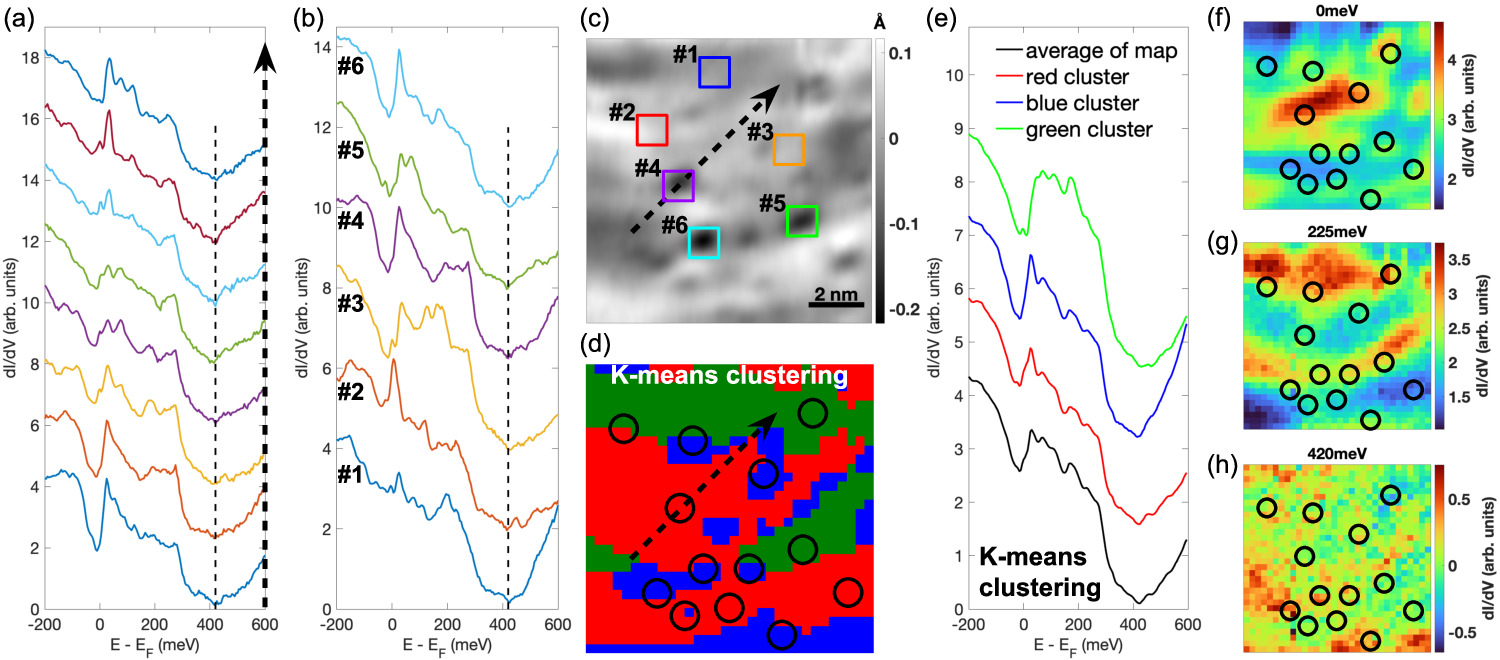}%
\caption{Density of states and differential-conductance on $EuSn_2As_2$. (a) Spatially resolved dI/dV spectra acquired along the dashed arrow in (c). The spectra are vertically shifted for clarity. (b) Averaged spectra collected over different regions of the area shown in (c). (c) Topographic image of the Sn surface of $EuSn_2As_2$. (d)
Results of K-means clustering analysis over the same area as in (c), identifying three distinct clusters. (e) Average local density
of states for each cluster, overlaid with the spatially averaged spectrum of the full area in (c). (f)-(h) Differential-conductance maps acquired over the area shown in (c) at different energies. The topograph and $dI/dV$ map were acquired at $V=-300\,\mathrm{mV}$ and $I_t=-400\,\mathrm{pA}$.}
\end{figure}

To elucidate the relationship between these electronic spatial variations and the underlying lattice, we perform spectroscopic imaging over the topographic region shown in Fig.2(c) and carry out K-means clustering to classify the local electronic signatures \cite{kavai2021inhomogeneous, kodinariya2013review, pedregosa2011scikit}. The K-means analysis identifies three dominant clusters [Fig.2(d)], with their cluster-averaged spectra shown in Fig.2(e) (Also see Supplemental Material Fig.~S4 for additional maps carried out on other parts of the sample).

The spatially resolved spectra and clusters reveal some spectral energy shifts and variations in the conductance near $200 \,\mathrm{meV}$. Figs.2(f)-2(h) show plots of the conductance maps corresponding to the local DOS at the Fermi energy [gap at $E_F$; Fig.2(f)], $225 \,\mathrm{meV}$ [Band edge; Fig.2(g)], the $420 \,\mathrm{meV}$ [small gap; Fig.2(h)]. We see that map at $E_F$ and $225 \,\mathrm{meV}$ correlate with the red and green clusters, respectively, whereas the map at $420 \,\mathrm{meV}$ shows more uniformity with minimal spatial variations. These K-means clusters and conductance maps, however, do not show a one-to-one correspondence with the surface Sn vacancies, which are overlayed in Figs.2(d) and 2(f)-2(h) as black circles, rather extend far beyond the immediate vacancy sites. The vacancies instead hole-dope the system and shift the Fermi energy deeper into the occupied states. Indeed, STM spectra measured across distinct sample regions reveal relative energy shifts of the order of $50 \,\mathrm{meV}$ (see Supplemental Material Fig.~S5).

The emergence of intrinsic hole-doping behavior appears to be a ubiquitous feature across the family of Eu-based 122 systems. In $EuSn_2As_2$, as evidenced by ARPES, $E_F$ shows a rigid band shift of approximately $180 \,\mathrm{meV}$ attributed to intrinsic hole doping \cite{li2019dirac}. Similar manifestations of hole doping are also reported in $EuSn_2P_2$ \cite{pierantozzi2022evidence, gui2019new}, $EuIn_2As_2$ \cite{gong2022surface,sato2020signature}, $EuZn_2As_2$ \cite{sar2026topologically} and $EuCd_2As_2$ \cite{santos2023eucd}. Collectively, these results indicate that the intrinsic hole doping shifts the Fermi level into lower binding energy and introduces significant spatial nanoscale inhomogeneity of the electronic states \cite{gong2022surface,sar2026topologically}.

A direct comparison between our dI/dV spectra and reported ARPES band structures \cite{li2019dirac} reveals a clear correspondence between the $300\text{--}500 \,\mathrm{meV}$ STM gap and the spin-orbit-induced bulk ARPES gap, the latter of which hosts the topological Dirac surface states. This experimentally observed bulk gap is also consistent with DFT predictions of the spin-orbit-induced band inversion gap at $\Gamma$ [Figs. 1(b)-1(c)], although the DFT results require a rigid energy shift due to intrinsic hole doping. The suppressed DOS in the STM spectra observed within this spin-orbit gap ($\sim300\text{-}500 \,\mathrm{meV}$ range) is consistent with the linear dispersion of the Dirac cone seen in ARPES \cite{li2019dirac}. Within this spin-orbit gap, a more refined analysis reveals a secondary, $\sim50 \,\mathrm{meV}$ gap centered at $\sim420 \,\mathrm{meV}$, which matches the energetic position of the gapless Dirac point observed in pump-probe ARPES \cite{li2019dirac}. To elucidate the role of magnetism in these electronic features, we investigate the temperature and spatial evolution of the dI/dV spectra across the AFM transition ($T_N = 24 \,\mathrm{K}$).

The temperature dependence of the average tunneling spectra, recorded over a $\sim10 \: nm$ area both below and above $T_N$, are shown in Figs. 3(a)-3(b). The data were collected within a same overall field of view, ensuring minimal variations in the Sn vacancy concentration. The most striking feature is the collapse of the partial gap at the Fermi level upon warming the system above $T_N$. A similar temperature-induced closing is also observed for the smaller ($50 \,\mathrm{meV}$) gap residing near $420 \,\mathrm{meV}$ within the larger spin-orbit gap. These analogous transitions provide strong evidence for the magnetic origin of both spectral features.

\begin{figure}
 \includegraphics[width=\textwidth]{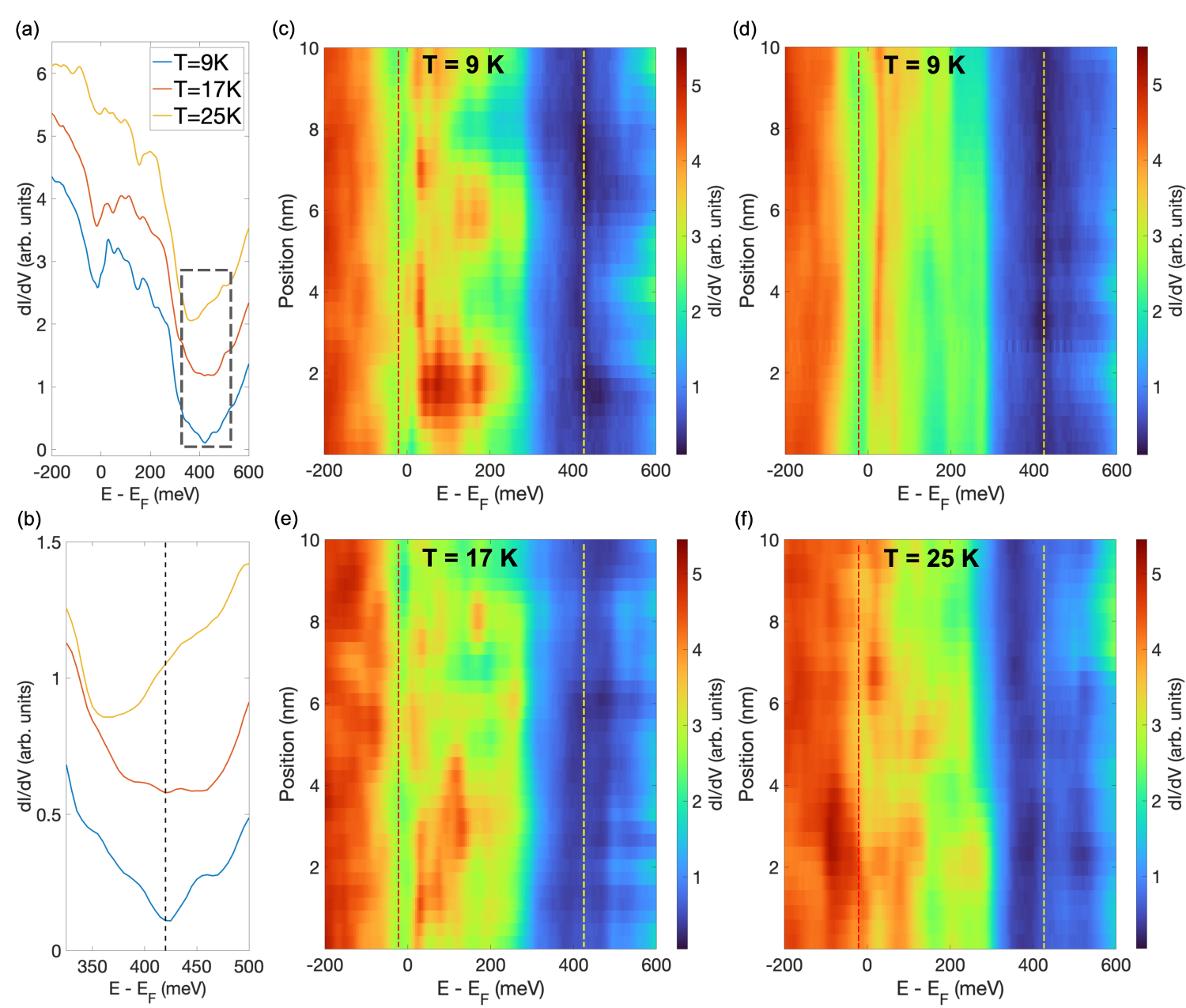}%
 \caption{Temperature-dependence of the electronic density of states across the AFM transition. (a) Evolution of the average dI/dV spectra across the magnetic phase transition temperature. (b) Enlarged view of the spectra in (a) highlighting the smaller ~50meV gap and its temperature evolution across the AFM transition. (c-f) Spatially resolved dI/dV spectra showing the spatial variations of the different spectral features across a 10nm line on the sample's surface at different temperatures. (c,d) are taken at 9K along a single K-means cluster (c) and across multiple clusters (d) highlighting homogeneity of the spatial variations. Temperature-dependent data were acquired at
$V=-300\,\mathrm{mV}$ and $I_t=-400\,\mathrm{pA}$.}
 \end{figure}

To resolve the spatial and temperature behaviors of these gaps, in Figs. 3(c)-3(f) we display dI/dV line cuts at the indicated temperatures across $T_N$. For $T = 9 \,\mathrm{K}$, we display two maps corresponding to a line cut spanning multiple K-means clusters [Fig.3(c)] and another through a single cluster [Fig.3(d)], showing the spatial variations across and within clusters. While the gaps manifest subtle nanoscale fluctuations on the order of a few meV, their systematic disappearance above $T_N$ confirms that the electronic gaps are intimately coupled to the long-range magnetic order. The spin-orbit gap, characterized by the low DOS between $300\text{-}500 \,\mathrm{meV}$ range does not show sensitivity to the AFM transition (as expected and observed in DFT and ARPES \cite{li2019dirac}). Based on Fig.3 and Supplemental Material (Figs.~S4-S5), the nano- and micro-scale variations in the energetic positions of the overall spectra can average out the temperature-dependent gaps at the global scale, which could be a reason to why these temperature-dependent gaps have not been resolved in the pump-probe ARPES measurements \cite{li2019dirac}. For example, the Dirac point observed in ARPES is located at $400 \,\mathrm{meV}$. The small gap position in our STM data varies from $320 \,\mathrm{meV}$ (see Supplemental Material Fig.~S5) to $420 \,\mathrm{meV}$ (Fig.3), demonstrating that a $50 \,\mathrm{meV}$ gap may not be easily resolved in a non-local measurement.

To further elucidate the origin of the observed spectral gaps, we perform DFT calculations of the bulk and surface electronic structures. Figure 4 presents the calculated surface states in the PM and AFM-b phases of $EuSn_2As_2$. The results show that the onset of AFM order induces a $100 \,\mathrm{meV}$ gap at the $\Gamma$ point near the experimental $E_F$ [marked with an arrow in Figs.4(b)-4(d)], mirroring the experimental gap observed in our STM data (Figs. 2 and 3). Furthermore, the calculations reveal a high DOS immediately above the gap, arising from a van Hove-like singularity at $\Gamma$. This feature directly corresponds to the gap-peak structure observed in the dI/dV spectra. We conclude that this gap originates from a combination of PM to AFM Brillouin zone folding and band hybridization inherent to the AFM phase.

To address the spectral gap observed in the $300\text{--}500 \,\mathrm{meV}$ range, characterized by a significantly suppressed DOS, we examine the calculated bulk band structure [Figs.1(b)-1(c)]. The DFT results reveal a robust spin-orbit-induced gap arising from the parity inversion of the Sn and As $p$-orbitals [Figs.1(b)-1(c)] at the $\Gamma$ point. The experimental characteristics of this gap, specifically its energy position, temperature independence, and the presence of sharp van Hove-like features at the band edges, are in excellent agreement with a topological gap origin. While the DFT-calculated gap magnitude is slightly smaller than the experimental value, such discrepancies are common. Indeed, recent studies have highlighted the sensitivity of this gap to the choice of exchange-correlation functionals and the Hubbard U parameter \cite{cuono2023ab}. Our experimental gap is highly consistent with the pump-probe ARPES dispersion and spin-orbit-gap \cite{li2019dirac}.

Finally, we address the $50 \,\mathrm{meV}$ gap that resides near $400 \,\mathrm{meV}$ and aligns with the gapless Dirac point seen in ARPES \cite{li2019dirac}. In the ideal AFM-b phase of $EuSn_2As_2$, the vertical mirror symmetry $M_y \:$(perpendicular to the magnetic moments) acts as a protector of the (001) surface Dirac states, potentially preserving a gapless crossing even in the absence of time-reversal symmetry $\mathcal{T}$ \cite{essin2009magnetoelectric, li2010dynamical, liu2020robust}. However, our observation of a $50 \,\mathrm{meV}$ gap opening below $T_N$ suggests a mechanism of symmetry-breaking possibly driven by the $(2.2 \pm 0.2)\%$ Sn vacancies. While global time-reversal symmetry $\mathcal{T}$ provides robust protection in the PM phase, the low-temperature magnetic order relies on the more fragile crystalline mirror $M_y$. The random distribution of Sn vacancies introduces local parity-breaking potentials that dismantle this mirror protection \cite{jing2026strain}, allowing the magnetic exchange field to induce a mass term at the Dirac point. This transition signifies a departure from the mirror-protected regime toward an axion insulator state, where the inversion-protected $Z_4$ topology is manifested through a gapped surface spectrum. In a very recent work on SnTe \cite{jing2026strain}, it was shown that intrinsic atomic defects locally and randomly break the mirror symmetry and, as a result, open a band gap in the topological edge states. This was also supported by theoretical simulations \cite{jing2026strain, liu2014spin}. The depth of the opened band gap depends on the extent to which mirror symmetry is broken.

\begin{figure}
 \includegraphics[width=\textwidth]{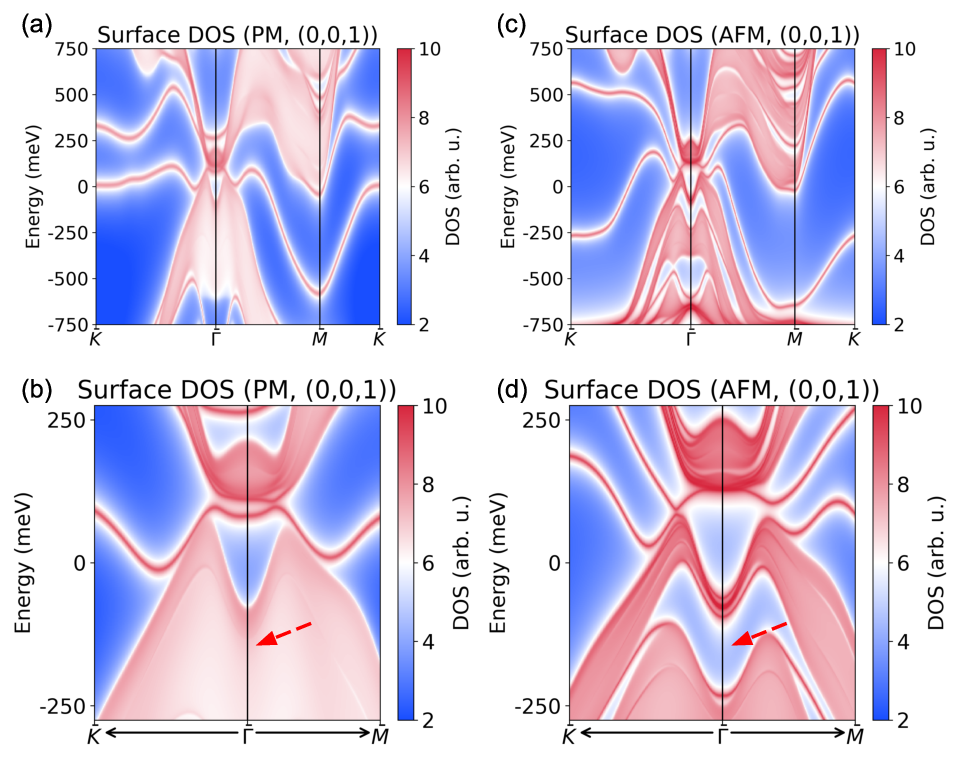}%
 \caption{Surface electronic structure calculations. (a,c) Calculated surface spectral function for the (001) surface in the PM (a) and AFM (b) phases. (b,d) Enlarged view near the Fermi energy corresponding to (a,c). The arrows in (b,d) correspond to the gap opening due to AFM folding and hybridization.}
 \end{figure}

We further consider whether this gap could arise from Brillouin zone (BZ) folding and band hybridization, analogous to the feature observed near the Fermi level. However, a comparison with ARPES data reveals a single Dirac cone with no evidence of such band splitting or folding within the spin-orbit induced gap. Theoretically, while the bulk bands must undergo Brillouin zone folding along the c-axis due to the magnetic unit cell doubling, the topological surface states are defined by in-plane momenta ($k_x$, $k_y$) and are localized at the vacuum interface where out-of-plane translational symmetry is already broken, making conventional BZ folding along c an unlikely origin for the $50 \,\mathrm{meV}$ gap. The absence of shadow bands or Dirac cone splitting in ARPES is consistent with this picture, suggesting the $400 \,\mathrm{meV}$ feature is more likely a mass gap induced by the local lifting of symmetry protection.

Our DFT calculations [Figs.4(a)-4(b)], however, present a more complex scenario. While the calculations confirm the presence of sharp, well-defined topological surface states, they deviate from the idealized topological insulator picture. Specifically, the DFT surface spectral function shows that bulk states near the $M$ point of the bulk Brillouin zone overlap in energy with the band-inversion gap at $\Gamma$ [Figs.1(b)-1(c)]. This overlap suggests the absence of a global bulk band gap, implying that the topological surface states may couple to bulk conduction states near the $M$ point. These results indicate that while the band-inversion topology is unambiguously present, the dispersion of the surface states may be highly sensitive to the local surface potential and hybridization with these overlapping bulk states.

Our STM dI/dV spectra, characterized by sharp band edges and a dramatically suppressed DOS in the $300\text{-}500 \,\mathrm{meV}$ range, suggests a lack of bulk states in this energy window, challenging the DFT prediction of $M$-point overlap. Furthermore, the idealized X-like Dirac cone seen in ARPES similarly points toward a clean gap. Note that ARPES does not explicitly probe the $M$-point. Based on this limited experimental evidence, the spectral feature observed in STM is most consistently linked to a magnetic exchange-driven mass gap of the Dirac point. Nevertheless, we cannot entirely rule out contributions from bulk AFM-induced hybridization: our DFT calculations indicate that bulk states near the $M$-point overlap in energy with the band-inversion gap (Fig. 4), potentially allowing coupling between bulk and surface states across the magnetic transition that could contribute to the observed gap.

\section{Summary}
In summary, our temperature dependent STM measurements on $EuSn_2As_2$ demonstrate a significant reconstruction of the electronic states across the AFM transition. We observe two distinct magnetically driven features: a $100 \,\mathrm{meV}$ gap near the Fermi level and a $50 \,\mathrm{meV}$ gap localized at the topological Dirac crossing. While DFT calculations suggest that the gap near $E_F$ may be primarily driven by Brillouin zone folding and band hybridization, the gap at the Dirac point occurs in a region where ARPES observes a single, unfolded Dirac cone. This suggests that the magnetic exchange field couples effectively to the topological states, likely enabled by the high concentration of intrinsic Sn vacancies that lift the crystalline mirror protections. While we cannot entirely exclude complex band hybridization effects for the $400 \,\mathrm{meV}$ feature, the coincidence of the gap with the Dirac crossing provides strong evidence for a magnetic gapping of the surface states. Our results highlight the importance of local probes in resolving the topological properties of magnetic insulators and establish $EuSn_2As_2$ as a potential platform for exploring the interplay between magnetism and topology.

\section{Methods}
\subsection{Sample Synthesis}
Single crystals of $EuSn_2As_2$ were grown using a Sn self-flux. High-purity elements (Eu $99.99\%$, Sn $99.99\%$, As $99.999\%$) taken in the atomic ratio Eu:Sn:As = $1.1:25:3$ were placed in an alumina $Al_2O_3$ crucible and sealed under high vacuum in a quartz ampoule. The ampoule was slowly heated to $1100\,^{\circ}\mathrm{C}$ and held at this temperature for $24\,\mathrm{h}$ to ensure complete homogenization of the melt. It was then slowly cooled to $700\,^{\circ}\mathrm{C}$ at a rate of $2\,^{\circ}\mathrm{C}/\mathrm{h}$. Upon reaching $700\,^{\circ}\mathrm{C}$, the ampoule was inverted and centrifuged to separate the excess molten tin flux from the crystals. Well-formed, shiny plate-like single crystals were obtained. The crystals remained stable in air over extended periods without observable degradation. The chemical composition and homogeneity were examined by energy-dispersive X-ray spectroscopy (EDS). The results confirmed good compositional homogeneity and a stoichiometry consistent with the chemical formula $EuSn_2As_2$.

\subsection{Magnetic and electrical transport measurements}
Magnetic measurements were performed in the temperature range $1.72$--$300\,\mathrm{K}$ and in magnetic fields up to $7\,\mathrm{T}$ using a Quantum Design MPMS-XL magnetometer. The electrical resistivity was measured in the temperature interval from $2$ to $300\,\mathrm{K}$ employing a Quantum Design PPMS-9 platform. Electrical contacts were made using $50\,\mu\mathrm{m}$ thick silver wires attached to single-crystalline specimens with silver epoxy paste.

\subsection{DFT Calculations}
First-principles calculations were performed within density functional theory (DFT) as implemented in the Vienna Ab initio Simulation Package (VASP) \cite{kresse1996efficient}, employing the projector augmented-wave (PAW) method to describe the ion--electron interaction. The exchange-correlation energy was treated within the generalized gradient approximation (GGA) in the parametrization of Perdew, Burke, and Ernzerhof (PBE) \cite{perdew1996generalized}. The Kohn-Sham
wave functions were expanded in a plane-wave basis set with a kinetic energy cutoff of $360\,\mathrm{eV}$. Brillouin zone integrations were performed on a $\Gamma$-centered Monkhorst--Pack $k$-point mesh of $11\times 11\times 3$. Strong on-site Coulomb interactions of the Eu $4f$ states were accounted for via the DFT+$U$ scheme with an effective Hubbard parameter of $U = 4.25\,\mathrm{eV}$ applied to the Eu $4f$ states. 

Maximally localized Wannier functions (MLWFs) were constructed using the Wannier90 code \cite{mostofi2014updated}, interfaced with VASP. The initial projections were chosen as the Eu $5s$ and $4f$ orbitals, the Sn $5s$ and $5p$ orbitals, and the As $4p$ orbitals. The disentanglement and frozen energy windows were chosen to faithfully reproduce the first-principles band structure across the relevant low-energy manifold. The resulting tight-binding Hamiltonian was subsequently used to investigate the surface electronic structure of the (001) termination via iterative Green's function methods as implemented in the
WannierTools package \cite{wu2018wanniertools}.

\subsection{STM/STS Measurements}
Samples were cleaved \textit{in situ} under ultrahigh vacuum at room temperature by knocking off an aluminum post. Immediately after cleaving, the samples were transferred into the STM and positioned next to a Cu(111) crystal used for PtIr tip preparation prior to each experiment. The sample and tip were subsequently cooled to the desired measurement temperature. Before tip preparation, the Cu(111) surface was cleaned by repeated cycles of sputtering and annealing. STM topographs were acquired in constant-current mode, while differential conductance
($dI/dV$) spectra were measured using a lock-in amplifier with a reference frequency of $0.921\,\mathrm{kHz}$.

\begin{acknowledgments}
P.A. acknowledges funding from the U.S. National Science Foundation (NSF) under award No. DMR-2406686. D.K. and T.R. were supported by the National Science Centre (Poland) under research grant 2021/41/B/ST3/01141.
\end{acknowledgments}

\bibliography{apssamp}

\end{document}